\documentclass[3p,times,twocolumn]{elsarticle}

\usepackage{ecrc}
\usepackage{amssymb}
\usepackage{amsmath}
\usepackage{graphicx}
\usepackage{epsfig,latexsym}
\usepackage{slashed}
\usepackage{simplewick}
\usepackage{subcaption}
\usepackage{nicefrac}

\def\be{\begin{eqnarray*}}
\def\ee{\end{eqnarray*}}
\def\beq{\begin{eqnarray}}
\def\eeq{\end{eqnarray}}
\def\bem{\begin{multline}}
\def\eem{\end{multline}}

\def\vac{\text{vac}}

\def\loss{\text{loss}}

\def\cut{\text{cut}}

\def\dd{\text{d}}

\newcommand{\onehalf}{{\nicefrac{1}{2}}}
\newcommand{\threehalfs}{{\nicefrac{3}{2}}}

\newcommand{\onethird}{{\nicefrac{1}{3}}}

\def\x{\boldsymbol{x}}

\volume{00}

\firstpage{1}

\journalname{Nuclear and Particle Physics Proceedings}

\runauth{}

\jid{nppp}

\jnltitlelogo{Nuclear and Particle Physics Proceedings}

\usepackage{amssymb}

\usepackage[figuresright]{rotating}

\begin{document}

\begin{frontmatter}



\dochead{}

\title{Parton energy loss in QCD matter}


\author{Konrad Tywoniuk}

\address{Theoretical Physics Department, CERN, Geneva, Switzerland}

\begin{abstract}
QCD jets, produced copiously in heavy-ion collisions at LHC and also at RHIC, serve as probes of the dynamics of the quark-gluon plasma (QGP). Jet fragmentation in the medium is interesting in its own right and, in order to extract pertinent information about the QGP, it has to be well understood. We present a brief overview of the physics involved and argue that jet substructure observables provide new opportunities for understanding the nature of the modifications. 
\end{abstract}


\end{frontmatter}


\section{Introduction}
\label{sec:intro}

The study of perturbative probes of the quark-gluon plasma, and QCD jets in particular, is currently in its golden age with the development of jet reconstruction techniques for heavy-ion collisions at LHC and RHIC, see e.g. \cite{Muller:2012zq,Spousta:2013aaa,Armesto:2015ioy} and these proceedings. These measurements provide in many ways a more rigorous connection between experimental measurements and theory or Monte-Carlo studies because of the implicit  resummation of collinear divergences. On the other hand, successful jet reconstruction in the extreme environment of heavy-ion collisions is challenging and comparisons between models and data should be done with care \cite{Cacciari:2010te,Cacciari:2011tm}.

Until recently most studies, both experimental and phenomenological, dealt with jet and di-jet rates as well as inclusive properties of jets, fragmentation functions and jet shapes, and measurements of large-angle energy flow around jets.  However, novel measurements of jet substructures in nuclear collisions \cite{CMS:2016jys} have recently invigorated the discussion and opened new possibilities for measuring and understanding medium modifications of jets.

The alleys of recent progress can predominantly be categorised according to two chief aspects of in-medium jet physics. Firstly, the propagation of a single colour charge in the medium and, secondly, the generalisation to multiple charges accounting for possible interference effects. We will review the former aspects in Sec.~\ref{sec:radiative} and the latter in Sec.~\ref{sec:coherence}. We will also discuss the application of the medium modifications on the level of jet substructure measurements in Sec.~\ref{sec:substructure}. This discussion is in no way meant to be exhaustive but will immediately illustrate the importance of whether sub-jets are treated as independent or coherent. Jet substructure provides therefore a new handle on the dynamics that can help pinpoint the microscopic processes underlying the measured modifications. 

The choice of focus here is of course a biased selection, and not all recent progress in the field can be covered. A very interesting topic which deserves further study is the back-reaction of medium dynamics to the propagation of the jet, see e.g. \cite{Zapp:2012ak,Wang:2013cia,He:2015pra,Casalderrey-Solana:2016jvj}. While these aspects certainly are important for quantitative comparisons to experimental data, we currently have not much to say about their qualitative features.
We summarise briefly in Sec.~\ref{sec:conclusions}.

\section{Radiative parton energy loss}
\label{sec:radiative}

A single hard parton traversing a coloured medium undergo successive elastic interactions which modify their kinematics, mainly leading to the transverse momentum broadening $\langle k_\perp^2 \rangle = \hat q L$, characterised by the parameter $\hat q$ in a medium of length $L$. The most efficient energy degradation mechanism is therefore realised through an enhanced rate of splitting. Assuming multiple soft scattering, the spectrum of induced quanta with energy $\omega$ radiated off a hard gluon is strongly cut-off at a characteristic energy $\omega_c \equiv \hat q L^/2$ and reads \cite{Baier:1996kr,Baier:1996sk,Zakharov:1996fv,Zakharov:1997uu}
\beq
\label{eq:BDMPSspectrum}
\omega\frac{\dd N_\text{\tiny BDMPS}}{\dd \omega} = \bar \alpha\left\{ \begin{array}{lc} \sqrt{\frac{\omega_c}{2 \omega}} & \omega  < \omega_c \\ \frac{1}{12} \left(\frac{\omega_c}{\omega} \right)^2 & \omega > \omega_c \end{array} \right.  \,,
\eeq
where $\bar \alpha \equiv 2\alpha_s N_c/\pi$. For further details and refinements, see e.g. \cite{Mehtar-Tani:2013pia,Blaizot:2015lma}.
The behaviour in the soft sector is characteristic of the Landau-Pomeranchuk-Migdal (LPM) interference between scattering centres and arises because the formation time of the gluon scales as $t_\text{f} = \sqrt{\omega/\hat q}$. One can also find a compact analytical expression for uncorrelated scatterings, the so-called ``first order in opacity'' spectrum \cite{Gyulassy:2000er,Wiedemann:2000za}; in this case, the LPM effect suppresses the hard sector.

The parameter $\omega_c$ determines the energy of gluons that have been broadened along the whole medium length and are emitted at the minimal angle $(\hat q L^3)^{-\onehalf}$. It is also controls the mean energy loss $\langle \Delta E \rangle \sim \hat q L^2 $. These emissions are rare $\mathcal{O}(\alpha_s)$, though. However, the energy scale $\omega_s = \bar \alpha^2\omega_c$ determines the regime when we have to take into account multiple branchings, i.e.  $\int_{\omega_s} \dd \omega \,\dd N_\text{\tiny BDMPS}/\dd \omega > 1$. Since their formation times is shorter than the medium length, a cascading process takes place which transports these gluons to large angles, $\theta > \bar \alpha^{-2}(\hat q L^3)^{-\onehalf}$. When we reconstruct the energy of the leading parton in a cone, this effect is responsible for sizeable energy leakage \cite{Blaizot:2013hx,Blaizot:2014ula,Blaizot:2014rla,Kurkela:2014tla}.

To get a clearer picture, let us put some numbers on these equations. For $L=4$ fm, $\hat q = 1$ GeV$^2$/fm and $\bar \alpha = 0.3$, we find $\omega_c = 80$ GeV and $\omega_s =7$ Gev. For this energy range, the corresponding range of emission angles, estimated from momentum broadening as $\theta \sim \sqrt{\hat q L}/\omega$, yields $0.025 < \theta_\text{\tiny BDMPS} < 0.28 $. For a jet reconstructed in a cone of $R=0.3$, this typical choice of medium parameters indicate that rare and hard BDMPS emissions populate the in-cone jet distribution while multiple branching transport energy out-of-cone. The details of this soft cascade has been studied in quite some detail and its connection with the physics of thermalisation has been highlighted \cite{Iancu:2015uja}. We will come back to this insight in Sec.~\ref{sec:substructure}.

As a reminder, we note that the soft emissions can be resummed into a probability distribution, called the quenching weight (QW), of losing a finite amount of energy \cite{Baier:2001yt,Salgado:2003gb,Baier:2006fr}. Taken the form of the spectrum in the first line of Eq.~(\ref{eq:BDMPSspectrum}), this distribution becomes
\beq
D_\text{\tiny QW} (\epsilon) = \sqrt{\frac{\omega_s}{\epsilon^3 }} \exp\left[-\frac{\pi \omega_s}{\epsilon} \right] \,,
\eeq
where a more realistic form can be tabulated \cite{Salgado:2003gb}. It can relate the jet spectrum in the presence of a medium to that in vacuum, $\dd N_{\text{jet}(0)}/\dd p_T^2 $, as
\beq
\label{eq:QuenchingFactor}
\frac{\dd N_{\text{jet}}}{\dd p_T^2} = \int_0^\infty \dd \epsilon \, D_\text{\tiny QW}(\epsilon) \frac{\dd N_{\text{jet}(0)}(p_T+\epsilon)}{\dd p_T^2 }\,.
\eeq
This allows to calculate the quenching factor $Q_\text{\tiny QW}(p_T)$ as the ratio of medium to vacuum spectra.

While the physics of transverse momentum broadening and radiative energy loss has been known for a while, recently progress has been made toward understanding their respective radiative corrections \cite{Liou:2013qya,Blaizot:2014bha,Iancu:2014kga}. Usually, one assumes that the interactions with the medium are quasi-instantaneous (with respect to the relevant timescales). However, allowing for short-lived, and thus soft, fluctuations one finds corrections which can most naturally be recast as corrections to the medium parameter $\hat q$. For instance, the first double-logarithmic correction reads
\beq
\Delta\hat q \simeq \frac{\alpha_s N_c}{2 \pi} \hat q \ln^2\frac{L}{l_0} \,,
\eeq
where the shortest timescale $l_0$ is some cut-off scale. The inclusion of these fluctuations to all orders leads to a renormalisation equation that accounts for a tower of fluctuations, ordered in formation time, and takes one from the value of $\hat q (l_0)$, i.e. describing the microscopic properties of the medium at scale $l_0$, to $\hat q(L)$, which includes the contribution from additional fluctuations in the medium. For a large medium, $\hat q (L) \propto L^\gamma$ where the anomalous dimension $\gamma = 2\sqrt{\bar \alpha}$ \cite{Blaizot:2014bha}. This novel relation affects how both the average transverse momentum broadening and energy loss scale with the size of the medium.

\section{Interference in multi-gluon processes}
\label{sec:coherence}

The ``running'' of $\hat q$ is an example of a resummation of fluctuations in the medium that overlap. In this particular situation, the fluctuations are strongly ordered and can easily be resummed. However, one could worry that in other situations, multiple fluctuations that interfere which each other would arise and thus ruin the probabilistic picture of independent emissions that underlies much of the discussion in the previous section. Besides, as known from jet physics in vacuum, these corrections give crucial input to Monte-Carlo shower generators of the fragmentation process and would serve for the same purpose for dedicated generators of jets in heavy-ion collisions.

The two-gluon rate in a dense medium was calculated in a series of noteworthy works \cite{Arnold:2015qya,Arnold:2016kek,Arnold:2016mth,Arnold:2016jnq}. They provided an independent confirmation of the double-logarithmic contributions discussed above. For most configurations the corrections to the probabilistic picture were small except whenever the gluon energies were strongly separated, i.e. one gluon being much softer than the other. Strikingly, in this case the found corrections were negative implying a reduced rate. This can be interpreted as an interference effect owing to the fact that, from the viewpoint of the shortest-lived fluctuation, the parent parton and the other, relatively long-lived fluctuation cannot be resolved \cite{Arnold:2016kek}. Physically this means that the shortest fluctuation can only be emitted off the total colour charge and not by each of the legs independently.

This striking result connects the physics of multiple medium-induced emission to the physics of jet fragmentation and modification in the medium. However, there are several subtle differences between the two cases. Firstly, splittings induced by the medium are not collinear divergent in contrast to vacuum radiation. Secondly, their formation is similar to their decoherence time, i.e. the time when a typical medium fluctuation can resolve it from the parent, see for a discussion on this point. These timescales can possibly differ a lot for vacuum radiation and we will come back to two cases below.

\begin{figure*}[t]
\centering
\begin{subfigure}[t]{0.5\textwidth}
\centering
\includegraphics[width=0.9\textwidth]{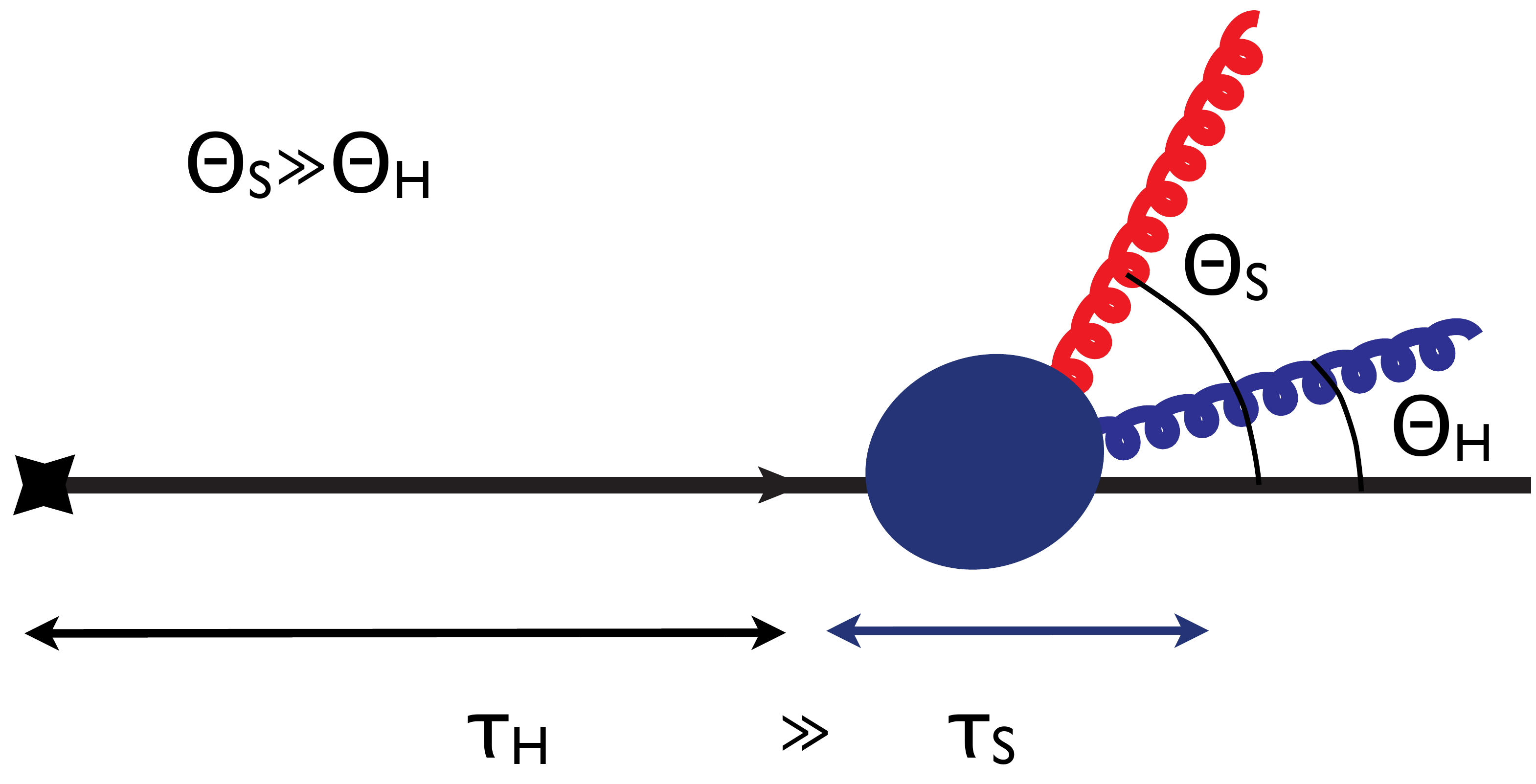}
\caption{}
\label{fig:CohEnLossa}
\end{subfigure}%
~
\begin{subfigure}[t]{0.5\textwidth}
\centering
\includegraphics[width=0.81\textwidth]{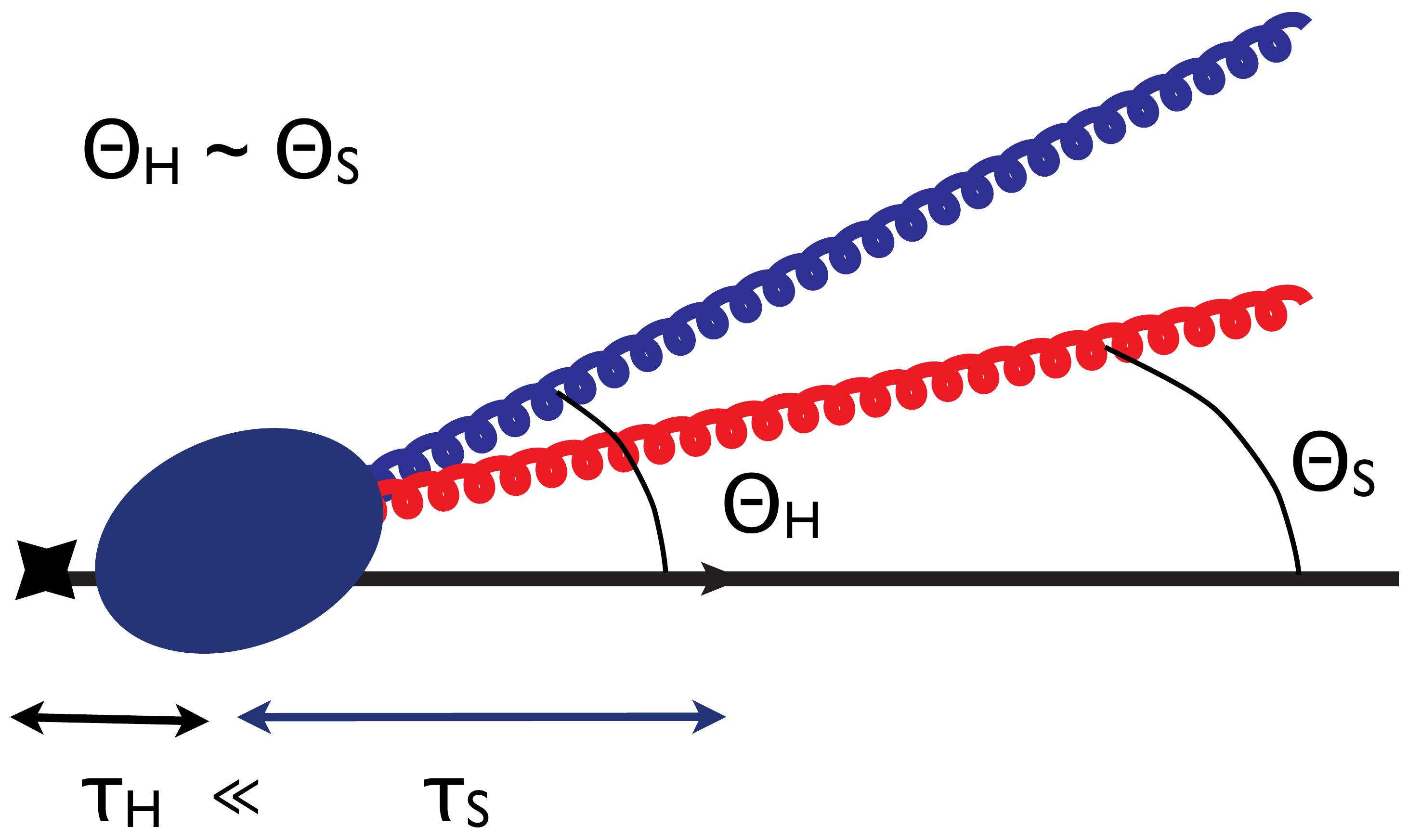}
\caption{}
\label{fig:CohEnLossb}
\end{subfigure}
\caption{Sketch of the two kinematic limits of the double emission rate calculated in \cite{Casalderrey-Solana:2015bww}. In both panels the hard gluon is blue while the soft gluon is red and the blob represents all possible placements of the in-medium exchange. (a) The collinear limit, left panel: the angle of emission of the hard gluon is very small and its formation time is long compared to the soft gluon formation time. (b) The soft limit, right panel: in this limit the formation time of the hard gluon is very short compared to the soft gluon one and the angles of emission of both gluons are comparable. Figures taken from \cite{Casalderrey-Solana:2015bww}.}
\label{fig:CohEnLoss}
\end{figure*}
In order to shed more light on these issues, one should consider the full two-gluon spectrum, differential in both energies and angles. While the full splitting function was first calculated in \cite{Fickinger:2013xwa} at first order in opacity, two limits of the spectrum, relevant for jet fragmentation in medium, were meticulously analysed \cite{Casalderrey-Solana:2015bww}. For simplification, one of the gluons was treated as ``hard'', i.e. its transverse momentum is much bigger than the medium kick, while the other not. Let us spend some time explaining these limits separately. These are illustrated in Fig.~\ref{fig:CohEnLoss}.

In the first limit, see Fig.~\ref{fig:CohEnLossa}, the formation time of the hard gluon is much longer than the formation time of the soft one. This is denoted the ``collinear limit'' since the hard gluon is emitted very close in angle to the parent parton. In fact, due to angular ordering the soft gluon is formally only radiated off the parent parton in the vacuum.\footnote{Soft emissions can only be emitted within a cone determined by the emitter. In the collinear limit, this cone shrinks to zero.} Nevertheless, in a large medium the two colour sources will ultimately be resolved and permitted to radiate. After this particular time one therefore finds an additional contribution to the spectrum, namely that of an emission spectrum off an on-shell colour current (Gunion-Bertsch spectrum). The timescale where the positive contribution to the rate sets in is simply the formation time of the hard gluon. Hence, the decoherence time is equal to the formation time or, in other words, the hard gluon gets resolved immediately after emission.

In the second limit, see Fig.~\ref{fig:CohEnLossb}, the formation time ordering is reversed. This happens whenever the energy of the soft gluon is small. In this case, the physical picture is quite intuitive: the parent parton and the hard gluon form a dipole that interact and radiate in the medium. In fact, one recovers exactly the spectrum off a colour charged ``antenna'' that was initially calculated at first order in opacity \cite{MehtarTani:2010ma,MehtarTani:2011gf} and generalised to multiple scattering in \cite{MehtarTani:2011tz,MehtarTani:2012cy,CasalderreySolana:2011rz}. In the latter, general situation the interference effects are controlled by the so-called decoherence parameter
\beq
\label{eq:DecoherenceParameter}
\Delta_\text{decoh} = 1-e^{- (L/t_\text{d})^{3}} \,,
\eeq
where we identify the decoherence time $t_\text{d} =[12/(\hat q \theta_\text{\tiny H}^2)]^{\onethird}$, where $\theta_\text{\tiny H}$ is the emission angle of the hard gluon. For long decoherence times, $t_\text{d} > L$, the dipole is not resolved by the medium and radiates medium-induced radiation coherently as the total colour charge. Additionally, it can radiate (fragment) vacuum-like according to the rules of angular ordering. In the opposite case, $t_\text{d} \ll L$, the dipole de-coheres, i.e. both constituents become independent of one another. Note that in both cases the decoherence time is much larger than the formation time, $t_\text{f} \ll t_\text{d}$.

Further work is need to understand intermediate regimes. Nevertheless, to summarise this section, the effects of colour coherence have been firmly established by several calculations. This points to a simple organising principle put forward in \cite{CasalderreySolana:2012ef}. Rewriting the decoherence parameter (\ref{eq:DecoherenceParameter}) to highlight a characteristic decoherence angle $\theta_\text{d} = \sqrt{12/(\hat q L^3)}$, one argues that the medium only can modify jet substructures at large angles $\theta > \theta_\text{d}$. The resolved substructures, in particular the jet core, fragment internally as in the vacuum and lose energy independently of one another. A significant fraction of typical jets in heavy-ion collisions could remain completely unresolved by the medium however they are still affected by energy loss effects due to the total (quark/gluon) colour charge of the jet. Corrections to this picture also can also account for the gradual eradication of angular ordering of the jet constituents and lead to an enhancement of soft gluons radiated within the jet cone \cite{Mehtar-Tani:2014yea}. Nevertheless, a complete understanding of how jets form and interact in the medium is still missing.

\section{Jet substructure in medium}
\label{sec:substructure}

In order to gain further insight into the mechanisms at play, and also encouraged by recent experimental measurements, it is natural to consider jet substructure observables. A particularly clear procedure, called ``SoftDrop''\footnote{Whenever $\beta = 0$, SoftDrop is equivalent to the modified MassDrop procedure \cite{Dasgupta:2013ihk}.} \cite{Larkoski:2014wba,Larkoski:2015lea}, selects a pair of subjets, starting from a maximal angular separation at the jet cone size $R$, that satisfies the criterion
\beq
\label{eq:SoftDrop}
z > z_\cut \theta^\beta \,,
\eeq
where $z\equiv \min({p_{T1},p_{T2}})/(p_{T1} + p_{T2})$, $p_{T1(2)}$ is the subjet energy and $\theta$ their angular separation. Candidates that do not satisfy the condition (\ref{eq:SoftDrop}) are discarded or ``groomed''. This procedure therefore corresponds to clustering all jet constituents into an angular ordered tree and look for the first ``hard'' branching, according to (\ref{eq:SoftDrop}). It is also worth keeping in mind that the procedure can be made to terminate at some minimal resolution angle $R_0$. Typical values chosen for the experimental analyses are $z_\cut =0.1$, $\beta = 0$ and $R=0.4$, $R_0=01$.

In vacuum, the ``hard'' branching is inherently sensitive to the fundamental splitting function, which for gluon-gluon splitting reads $\mathcal{P}^\vac(z,\theta) = \bar \alpha P(z)/\theta$ where $P(z)$ is the relevant Altarelli-Parisi splitting function (stripped of its colour factor). However, given it's collinear divergence $\sim\bar \alpha \ln\big(R/R_0 \big)$ ($\beta =0$) we have to resum multiple emissions into the relevant Sudakov form factor. Physically, this means taking into account all the groomed emissions for $R\gg R_0$. We can then, for instance, define the probability to split to two sub-jets with momentum fraction $z_g$ as
\beq
\label{eq:SplittingProbVacuum}
\textsl{p}(z_g) = \int_{0}^R \dd \theta\,  \Delta(\theta) \mathcal{P}^\vac(z_g,\theta) \Theta_\cut(z_g,\theta) \,,
\eeq 
where $R_0\to0$ and the step-function in (\ref{eq:SplittingProbVacuum}) embodies the condition in Eq.~(\ref{eq:SoftDrop}), for details see \cite{Larkoski:2014wba,Larkoski:2015lea,Mehtar-Tani:2016aco}. The relevant Sudakov reads
\beq
\Delta(\theta) = \exp \left[-\!\!\int_\theta^R \!\!\dd \theta' \!\!\int_0^1 \!\!\dd z \, \mathcal{P}^\vac(z,\theta') \Theta_\cut(z,\theta') \right] \,,
\eeq
and is equivalent to the 1-jet rate, i.e. it is the probability of no splittings between the maximal angle $R$ and $\theta$. Given a resolution angle $R_0$, the probability of finding a pair that satisfies the SoftDrop condition, aka the two-pronged probability, is therefore $\mathbb{P}_{2\text{prong}} = 1-\Delta(R_0)$ \cite{Mehtar-Tani:2016aco}.
Strikingly, after the resummation the splitting {\it probability} becomes independent of $\bar \alpha$, thus not on the value of $\alpha_s$ nor the colour or {\it flavour} of the splitting, and exhibits the universal $1/z$-behaviour at small-$z$ for the $\beta = 0$ case \cite{Larkoski:2015lea}.

When considering the medium modifications of this observable, we are guided by the insight found in the previous sections that imply an approximate separation of two types of radiation: multiple, soft on large-angles and rare, hard emission in the jet cone \cite{Mehtar-Tani:2016aco}, see also \cite{Chien:2016led}. Hence, having to deal with two sub-jets we have to decide, according to some criterium, whether they lose energy coherently or independently. Secondly, for jets with $p_\text{T} = 100-200$ GeV our back of the envelope estimate shows that hard BDMPS radiation could be identified by the SoftDrop as actual jet substructures. The effect should be small $\mathcal{O}(\alpha_s)$ and care should be taken when aiming for a quantitative a description of the data. Nevertheless, let us come back to this exciting point later and currently focus on the first aspect, sub-jet coherence.

Due to energy loss effects the probability of the splitting is intimately related to the suppression of the spectrum itself. In order to simplify the discussion, let us consider two clearly defined scenarios and review their consequences, for more details see \cite{Mehtar-Tani:2016aco}. In the first scenario the whole jet, and therefore all its sub-jets, is unresolved by the medium. In the second scenario all sub-jets are resolved, thus independent. In order to study these scenarios we will make use of a probabilistic setup where energy loss (whether elastic or radiative) can affect any resolved sub-jet.

In the former, ``coherent'' case none of the inter-jet splittings are modified but the spectrum is overall suppressed because of energy loss, as given by Eq.~(\ref{eq:QuenchingFactor}). This implies that, in the absence of any other source of radiation, Eq.~(\ref{eq:SplittingProbVacuum}) holds. The proper way of adding a new radiative mechanism, namely in-cone BDMPS emissions, is on the level of probabilities. Hence, we have to reduce the vacuum probability in order to obtain a properly normalised total probability of radiation. After taking appropriate care of the angular restrictions (for instance, the introduction of a minimal resolution angle should further suppress the contribution of vacuum radiation) we should expect an enhancement of the splitting probability at small-$z$ because of the medium-induced bremsstrahlung that scales as $z^{-\threehalfs}$. This enhancement dies rapidly off with energy $\sim p_T^{-\onehalf}$, see Eq.~(\ref{eq:BDMPSspectrum}). In effect, the two-pronged probability $\mathbb{P}_{2\text{prong}}$ should be enhanced compared to the vacuum.

Taken at face value, this scenario illustrates that the SoftDrop procedure presents a unique possibility to measure directly medium-induced quanta rather than simply being sensitive to its general consequences, such as energy loss, etc.

The second scenario sketched above is more complicated. Let us first analyse the effects of energy loss for vacuum radiation in a limited angular range, $R \gtrsim R_0$. The splitting probability is now explicitly convoluted with the final-state jet spectrum, and reads ($\beta =0$)
\begin{align}
\label{eq:SplittingFunctionIncoherent}
&\frac{\dd N}{\dd p^2_T}\textsl{p}(z_g) = \bar \alpha \ln\frac{R}{R_0} \int_0^\infty\dd \epsilon \int_0^\epsilon \dd \epsilon' D_\text{\tiny QW}(\epsilon-\epsilon') D_\text{\tiny QW}(\epsilon') \nonumber\\
&\times \frac{p_T}{p_T+\epsilon} P\left(\frac{z_g p_T + \epsilon'}{p_T+ \epsilon} \right)\frac{\dd N_{(0)}(p_T+\epsilon)}{\dd p^2_T} \Theta(z_g-z_\cut) ,
\end{align}
for $z_g< 1/2$. This time the splitting function itself is directly affected by the fact that energy loss of the outgoing legs is independent. This can be seen by expanding the Altarelli-Parisi splitting function for $\epsilon,\epsilon' \ll p_T$ in the small-$z_g$ region where it reads
\beq
P\left(\frac{z_g p_T + \epsilon'}{p_T+ \epsilon} \right) \simeq \frac{1}{z_g}\left(1-\frac{\epsilon'}{z_g p_T} \right) \,.
\eeq
The characteristic energy-splitting variable can be seen to shift as $z_g \to z_g + z_\loss$, where $z_\loss \sim \omega_s/p_T$ from dimensional arguments, resulting in a flattening of the $z_g$-distribution. Furthermore, Eq.~(\ref{eq:SplittingFunctionIncoherent}) contains two quenching weights in contrast to only one in the ``coherent'' scenario. This signals for the first time the strong effects of energy loss when applied to incoherent substructures within the original jet.

In order to understand how to disentangle the  splitting probability in Eq.~(\ref{eq:SplittingFunctionIncoherent}), imagine a situation where most of the quenching the jet as a whole is taken by the most energetic leg (carrying momentum fraction $1-z_g$, for $z_g <1/2$). The jet spectrum on the left-hand side of Eq.~(\ref{eq:SplittingFunctionIncoherent}) is again given by (\ref{eq:QuenchingFactor}). The remaining quenching affects only the soft leg and can now be resummed into a modified Sudakov form factor that accounts for energy loss. This resummed quenching effect strongly suppresses the probability of two-pronged objects compared to the vacuum.

It becomes clear that adding the BDMPS spectrum on the level of probabilities complicates the situation further and it is not our goal here to present a definite answer. We could argue that the strong effects of incoherent energy loss strongly distorts the vacuum spectrum, thus being hard to reconcile with the trends observed in experimental data. A more realistic calculation should provide an interpolation between the two extreme scenarios discussed so far. Besides, effects of a soft background correlated with the jet, e.g. generated by back-reaction, could influence the interpretation of the result. 

Nevertheless, the potentially unique prospect of a (semi-)direct measurement of the medium-induced bremsstrahlung and its interplay with jet coherence in heavy-ion collisions motivate further investigations into this and related jet substructure observables.

\section{Conclusions \& outlook}
\label{sec:conclusions}

Jet physics in medium is currently witnessing notable advances from the theory side and enjoys a well of excellent experimental data that continues to push for further improvements. It is therefore pertinent to understand the process of jet fragmentation in a medium in great detail. Only then can we claim to extract reliable information about the properties of the medium.

In many cases, we can however completely neglect in-cone jet modifications with a suitable adjustment of medium parameters. Jet substructure measurements are a door-opener in this context since they demand a treatment of well-defined sub-jets. The guiding insights come from the analysis of both the fragmentation of soft medium-induced gluons and the study of interference effects of hard radiation. These new class of observables also allow to test and benchmark these insights against full-fledged Monte Carlo generators for jets in heavy-ion collisions, e.g. \cite{Casalderrey-Solana:2016jvj,Zapp:2012ak}. This promises a very fruitful synergy in the future.




\section*{Acknowledgements}
Thank you Y. Mehtar-Tani and J. Casalderrey-Solana for fruitful discussions. KT has been supported by a Marie Sk\l{}odowska-Curie Individual Fellowship of the European Commission's Horizon 2020 Programme under contract number 655279 ``ResolvedJetsHIC''.

\nocite{*}
\bibliographystyle{elsarticle-num}
\bibliography{jos}



\end{document}